\documentstyle[12pt]{article}
\newlength{\dinwidth}
\newlength{\dinmargin}
\def\bbbz{{\mathchoice {\hbox{$\sf\textstyle Z\kern-0.4em Z$}}
{\hbox{$\sf\textstyle Z\kern-0.4em Z$}}
{\hbox{$\sf\scriptstyle Z\kern-0.3em Z$}}
{\hbox{$\sf\scriptscriptstyle Z\kern-0.2em Z$}}}}
\def\be{\begin{equation}}
\def\ee{\end{equation}}
\def\ba{\begin{eqnarray}}
\def\ea{\end{eqnarray}}

\def\H{{\cal H}}
\def\L{\Lambda }
\def\half{\frac12}

\newcommand{\A}{{\cal A}}
\def\tilde#1{\widetilde{#1}}

\def\rho{\varrho}
\def\epsilon{\varepsilon}

\begin{document}
%
\newcommand{\skipp}[1]{\mbox{\hspace{#1 ex}}}
\newcommand{\vskipp}[1]{\vspace{#1 pt}}
\newcommand{\cit}[1]{$^{#1}$}
\newcommand{\fig}[6]{\setlength{\unitlength}{1mm}
                     \begin{figure}[tbh]
                     \begin{center}
                     \begin{picture}( #1 , #2 )
                     \special{#3}
                     \end{picture}
                     \end{center}
                     \vspace{#4}\baselineskip 10pt\begin{center}
                      \begin{quotation}
                     {\footnotesize Fig.\ #5\,.\ #6}
                      \end{quotation}
                     \end{center}\end{figure}
                     \baselineskip 13pt \vskipp{2}}
\newcommand{\tab}[3]{\vskipp{5}\baselineskip 10pt\begin{center}
                      \begin{quotation}
                     {\footnotesize Table #1\,.\ #2}
                     \vskipp{0} \end{quotation} \end{center}
                     \begin{table}[h]
                     \begin{minipage}{5truein}{}
                     \begin{center}
                     {\footnotesize #3}
                     \end{center}
                     \end{minipage}
                     \end{table}
                     \baselineskip 13pt\vskipp{2}}
 \newcommand{\Dirac}{\not\!\nabla}
 \newcommand{\gsim}{\raisebox{-3pt}{$\scriptstyle\stackrel{>}{\sim}$}}
 \renewcommand{\arraystretch}{1.5}
 \noindent{\tt DESY $92-126$ \hfill ISSN $0418-9833$}\\
 {\tt September 1992} \hfill hep-lat/9209021
 \mbox{}

 \renewcommand{\thefootnote}{{\protect\fnsymbol{footnote}}}

 \begin{center}
 \mbox{}
 \vspace{3cm}

 {\LARGE  Neural multigrid for gauge theories\\
          and other disordered systems} \\

 \vspace{1.5cm}
             M. B\"aker,
             T. Kalkreuter,
             G. Mack
             and M. Speh \\  \smallskip
            II. Institut f\"ur Theoretische Physik,
            Universit\"at Hamburg,\\
            Luruper Chaussee 149, 2000 Hamburg 50
 \vspace{1.5cm}
 \mbox{}

 \end{center}
 \begin{center}{\footnotesize ABSTRACT} \end{center}
\begin{quote}

   {\footnotesize
We present evidence that
multigrid works for wave equations in disordered systems, e.g. in
the presence of gauge fields, no matter how
strong the disorder, but one needs to introduce a ``neural computations''
point of view into large scale simulations: First, the system must
learn how to do the simulations efficiently, then do the simulation
(fast). The method can also
be used to provide smooth interpolation kernels which are needed in
multigrid Monte Carlo updates.
   }

\vspace{5pt}

\end{quote}
 \baselineskip 13pt
 \renewcommand{\thefootnote}{\protect\alph{footnote}}
 \vspace{5pt}
   There is a stochastic multigrid method and a deterministic one.      The
stochastic version is used to compute high dimensional integrals in
Euclidean quantum field theory or statistical mechanics by a Monte Carlo
method which uses updates at different length scales~\cite{1,2}. The
deterministic version~\cite{3,4} solves discretized partial
differential equations. One hopes to use both of them in simulations
of lattice QCD, for updating the gauge fields and for computing
fermion propagators in given gauge fields. In either case the aim is
to beat critical slowing down in nearly critical systems,
i.e. to maintain fast  convergence when long range correlations appear.

A crucial problem is how to define and exhibit smooth functions
in the disordered context, i.e. when translation symmetry is strongly
violated. We will present a method how to solve this problem.

We recommend to think of more general disordered systems than gauge
theories. This puts the core of the problem into sharper focus, and it
opens the way to other possible applications such as
%
%
low lying states of spin glasses,
the shape of a lightning,
waves on fractal lattices (with bond percolation),
localization of low lying electronic states in amorphous materials, etc.
\clearpage
 \addtocounter{section}{1}
 \setcounter{footnote}{0}
 \setcounter{equation}{0}
 \setcounter{figure}{0}
 \setcounter{table}{0}

 {\bf\noindent 1. The Multigrid}
\vskipp{5}

One starts from a problem on a given ``fundamental'' lattice $\L^0 $
of lattice spacing $a_0$. One introduces a sequence of lattices
$\L^1,\L^2,..., \L^N$ of increasing lattice spacings $a_j=L_b^ja_0$,
together with
interpolation operators $\A^j $ and restriction operators $C^j$
which map functions on coarser lattices into functions on
finer lattices, and vice versa.
Let $\H^j$ be the space of functions on lattice
$\L^j$. Then we need
\ba &\A^j&: \ \H^j \mapsto \H^{j-1}:
                         \ \ \mbox{interpolation operator} \\
                 &C^j&: \ \H^{j-1} \mapsto \H^j:
                         \ \ \mbox{restriction operator}
\ea
Given the interpolation operators, one can use them to define
restriction operators, The choice
 $C^j = \A^{j \ast }$ is made in ``variational coarsening''~\cite{2}.

Typically, we choose $L_b=2$, and a last layer $\L^N$ which consists of
a single point.
 \addtocounter{section}{1}
 \setcounter{equation}{0}
 \setcounter{figure}{0}
 \setcounter{table}{0}

\vskipp{12}
 {\noindent \bf
   2. The Basic Importance of Smoothness}

\vskipp{5}
{\noindent \em $2.1.$ Deterministic multigrid}

\vskipp{5}

One wants to solve a discretized elliptic differential equation on
$\L^0$
\be D_0\xi^0 =f^0 \ .  \label{De}
\ee
It might have arisen from an eigenvalue equation
$D_0\xi^0 = \epsilon \xi^0 $ by inverse iteration~\cite{5}.
                                                   If $D_0$ has a small
eigenvalue, then local relaxation algorithms suffer from critical
slowing down.

{\bf Basic observation} (in the ``or\-dered case''~\cite{3,4}):
After some (damped) relaxation sweeps on $\L^0$ one gets an approximate
solution $\tilde{\xi }^0 $ whose error
  $ e^0 = \xi^0 - \tilde{\xi }^0 $
is not necessarily small but is \it smooth \rm (on length scale
$a_0$). The unknown error $e^0$
satisfies the equation
\be D_0 e^0 = r^0 . \label{Dee} \ee
It involves the \it residual \rm
 $r^0 = f^0 -D_0\tilde{\xi }^0 $
which would be zero for an exact solution.

Given that $e^0$ is smooth, it can be obtained by smooth interpolation
of a suitable function $e^{1  } $ on $\Lambda^1$,
\be e^0 = \A^1 e^{1 }\ . \label {intp0}  \ee
That is, $e^0_z = \sum_{x \in \L^1} \A^1_{zx}e^{1  }_x $ with
$\A^1_{zx}$ which depends smoothly on $z$.

Now define a restriction operator $C^1$ such that interpolation followed
by restriction amounts to doing nothing, i.e. $C^1\A^1=1$.
(For instance   $C^1 =(\A^{1\ast }\A^1)^{-1}\A^{1\ast}$).
Then~(\ref{intp0}) can be inverted,
 $ e^{1  } = C^1 e^0 \ . $
Applying $C^1 $ to both sides of (\ref{Dee}) yields an equation for
$e^{1  }$,
\be D_1 e^{1  } = r^1 \ , \label{cgrideq}\ee
with
$ r^{1  } =  C^1r^0 $ \  (restricted residual)
 and $ D_1 = C^1D_0\A^1 $ (effective differential operator).
Given $e^{1  }$, one obtains $e^0$ from~(\ref{intp0}), and
$\tilde{\xi }^0 + e^0$ is an 
improved solution of~(\ref{De}).
Thus,\it the problem has been reduced to an equation on the lattice
$\L^1 $ which has fewer points. \rm
If necessary, one repeats the procedure, moving to $\L^2 $ etc. The
procedure stops, because an equation on a ``lattice'' $\L^N$ with only a
single point is easy to solve.

The iterated interpolation
 $ \A^{[0j]} \equiv \A^1 \A^2 ... \A^j$
                           from $\L^j$ to $\L^0$ should yield functions
on $\L^0 $ which are \it
           smooth on length scale $a_j$\rm , i.e. which change
little over a distance $a_j$ (in the ordered case).
For reasons of practicality, one must require that
 \be \A^j_{zx}= 0 \ \mbox{unless $z$ is near $ x$  .}  \ee
\indent {\bf Example:}
The optimal choice for the 1-dimensional Laplace equation is~\cite{4}
\be
\A_{xx} = 1 \quad , \A_{x\pm 1,x}= \half \quad , \mbox{ others} = 0 \ .
\ee
We chose to regard $\L^1 $ as a sublattice of $\L^0$ and set $a_0=1$.

This
     interpolation has the property that it
                   maps constant functions on $\L^1$ into constant
functions on $\L^0$. Constant functions are the lowest eigenmodes of the
Laplacian.
$D_1$ comes out proportional to a Laplacian again.

\vskipp{12}
{\noindent \em $2.2.$ Stochastic multigrid}
\vskipp{5}

The successful stochastic multigrid updating method~\cite{6,7}
for $O(N)$, $CP^N$ and $SU(N)~\times SU(N)$ spin models was described in
Wolff's lecture~\cite{8}. One uses updates of spins $s(z)$ at sites
$z \in \L^0$ of the form
\be s(z)= e^{i\lambda \A^{[0j]}_{zx}}s(z) \ee
where the matrix $\lambda $ is a generator of a group of transformations
 which can act on spins $s$, and where $\A^{[j]}_{zx}$ vanishes outside
 a neighborhood of diameter of order $a_j$ of $x$. One may assume
 \footnote{In most simulations, it was actually chosen at random.}
 $x$ to take values in a lattice $\L^j$.
 It is important, though, that
the supports of $\A^{[0j]}_{zx}$ in $z$ should overlap for adjacent $x$.

 This procedure eliminates critical slowing down in the spin models
                                                 almost completely,
 \it provided \rm on chooses $\A^j $ to be smooth in $z$ on length scale
 $\a_j$.

 For general models, a sufficiently high acceptance rate for
 nonlocal updates like (2.7) is necessary to eliminate critical
 slowing down. Pinn and Grabenstein show evidence~\cite{9} that
 smoothness alone is not in general enough to ensure this.
 Their criterion demands also nonappearance of mass terms.
                                                     In the spin models
 this is true.
 Pure gauge theories have no bare mass parameter. But an effective mass
is present in the
 following ``weak coupling''
          multigrid updating scheme for gauge fields in 4 dimensions,
 with gauge group $G$, cp.~\cite{9}.
 Suppose one updates only gauge fields  attached to links which
 point in one selected direction - call it the vertical direction.
 One can regard these variables as spins, and perform updatings like
 (2.7), with $\A $ that are smooth in an appropriate sense -
 cp. later. Only variables residing in the same horizontal 3-dimensional
 sublattice are actually coupled, therefore one effectively does
 updatings in  3-dimensional Higgs-models, with action
 $ -\beta s(z)\Delta^{\perp} s(z)$. The covariant Laplacian
 $\Delta^{\perp }$ depends on a 3-dimensional $G\times G$-gauge field
 that is determined by the gauge field variables of the model which are
 attached to horizontal links. The lowest eigenvalue of $\Delta^{\perp}$
 has dimension mass squared, and is strictly positive for generic
 gauge fields. It is determined by variables that are not updated and
 is therefore like a parameter.

 Experience with $\phi^4$-theory suggests that such a ``weak coupling"
 approach might nevertheless lead to a substantial acceleration in
 practise. This is under investigation.
 \addtocounter{section}{1}
 \setcounter{equation}{0}
 \setcounter{figure}{0}
 \setcounter{table}{0}

\vskipp{12}
 {\noindent \bf 3. Smoothness in Disordered Systems}
\vskipp{5}

{}From section 2 we learn that a successful multigrid scheme, whether
deterministic or stochastic, needs smooth interpolation kernels $\A $.
This raises the basic

{\bf Question:}\it What is a smooth function in the disordered
 situation, for instance in an external gauge field? \rm

 Naive smoothness of a function $\xi $ means that
 $
 \sum_{\mu } (\nabla_{\mu }\xi , \nabla_{\mu }\xi )\ll (\xi , \xi ) ,
 $
 where $\nabla_{\mu }$ are discretized ordinary derivatives.
 But this is not gauge covariant. A tentative remedy would be to
 take the covariant derivative $\nabla_{\mu }$. But
 \be
 \sum_{\mu } (\nabla_{\mu }\xi , \nabla_{\mu }\xi )
 = (\xi , -\Delta \xi ) \geq \epsilon_0 (\xi , \xi )\ . \ee
 The lowest eigenvalue $\epsilon_0$ of the negative
 covariant Laplacian $-\Delta $
 is a measure for the disorder of the gauge field. (It is positive and
 vanishes only for pure gauges.) By definition, it is not small for
 disordered gauge fields. Therefore there are no smooth functions in
 this case.

 Nevertheless there is an answer to the question, assuming
 a fundamental differential operator $D_0$ is specified by the problem.
 In the stochastic case, the Hamiltonian often provides
 $D_0$.

 {\bf Answer: }
 \it A function $\xi $ on $\Lambda^0$ is smooth on length
 scale $a$ when   \rm
 \be \| D_0\xi \|^2 \ll \|\xi \|^2 \ \mbox{ in units }a=1 \ . \ee

 We  found that  a deterministic multigrid which employs
 interpolation kernels
                    $\A^{[0j]}$ from $\L^j $ to the fundamental lattice
 $\L^0$ which are smooth in this sense, works for arbitrarily disordered
 gauge fields - see later.

 When there are no smooth functions in this sense at length scale $a_0$,
 then $D_0$ has no low eigenvalue, and there is no critical slowing
 down and no need for a multigrid.


 The above answer appears natural, and the
``projective multigrid'' of~\cite{10,11} is in its spirit.
 But there are subtleties, and
 there is the question of how to obtain  kernels $\A^{[0j]}_{zx}$
 which are smooth \it on length scale $a_j$.  \rm

 {\bf Problem:} \it One needs  approximate solutions of eigenvalue
 equations
 \be D_0 \A^{[0j]}_{zx} = \epsilon_0(x) \A^{[0j]}_{zx}\ . \label{EVA}
 \ee  \rm
 $x$ is fixed, and $D_0$ acts on $z$. Since $\A^{[0j]} $ is required to
 vanish for $z$ outside a neighbourhood of $x$, the  problem involves
 Dirichlet boundary conditions.

 For large $j$, $\A^{[0j]} $ will have a large support.
                                          If there is no degeneracy
 in the lowest eigenvalue, one can use inverse iteration combined
 with standard relaxation algorithms for the resulting inhomogeneous
 equation\footnote{Basically one computes an approximation to
 $D_0^{-n}\A^{[0j]}_{start}$.}.
           But this and other standard methods will
 suffer from critical slowing down again.

 Moreover, in the standard multigrid setup, one uses basic interpolation
 kernels $\A^j$ which interpolate from one grid $\L^{j}$ to the next
 finer one. In this case
  \be \A^{[0j]} = \A^1 \A^2 ... \A^j \ , \label{FACTOR} \ee
 and (\ref{EVA}) becomes a very complicated set of nonlinear conditions.

 Possible solutions are
 \begin{description}
 \item[{\rm (i)}] Replace (\ref{EVA}) by minimality of a cost functional
       (cp. later). Use neural algorithms to find kernels $\A^j $
       which minimize it.  This is still under study.
 \item[{\rm (ii)}] Give up factorization (3.4) and determine independent
       kernels $\A^{[0j]}$ as solutions of (\ref{EVA}) by
       multigrid iteration. This is done
             successively for $j=1,2,\dots$ One uses
       already determined kernels $\A^{[0k]}$ with $k<j$ for updating
       $\A^{[0j]}$. We found that this works very well - cp. later.
 \end{description}

 Method (ii) is not quite as efficient as
 standard multigrid methods (MG) for
 ordered systems because of the overhead for storing and computing the
 kernels. Assuming convergence as expected, algorithms compare as
 follows for a lattice $\L^0 $ of $L^d $ sites in $d$ dimensions.
 (The overhead is included):
\begin{center}
\begin{tabular}{|l|l|l|}\hline
  Algorithm             & work (flops)               &storage\ space\\ \hline
 local\ relaxation      & $ L^{d+z} (z\approx 2)$    & $L^d$         \\
 MG [``ordered'']        & $ L^d $                    & $L^d$         \\
 MG [``disordered'', (i)]& ???                  & $L^d$         \\
 MG [``disordered'', (ii)]& $L^d\ln^2 L$        & $L^d \ln L$ \\ \hline
\end{tabular}
\end{center}
 \addtocounter{section}{1}
 \setcounter{equation}{0}
 \setcounter{figure}{0}
 \setcounter{table}{0}

\vskipp{12}
 {\noindent \bf 4. Criteria for Optimality}
\vskipp{5}

 We consider iterative solution of a discretized partial
differential equation~(2.1).
Any iteration amounts to updating steps of the
 form
 \be \tilde{\xi}^0 \mapsto \tilde{\xi}^{ 0\prime } = \rho\
\tilde{\xi}^0 + \sigma\ f^0 \ .
                              \label{iterlin}                    \ee
 $\rho$ is called the iteration matrix, and $\sigma =(1-\rho )D_0^{-1}.$
 The convergence  is governed by the norm $\|\rho \|$ of the
 iteration matrix. The iteration converges if $\|\rho \|< 1$, and
 its relaxation time is
 \be \tau \leq -\frac{1}{\ln \|\rho \|} \ . \ee
 \it Parameters \rm in the algorithm - such as interpolation kernels
 $\A^j_{zx}$, restriction operators $C^j_{xz}$ and effective
 differential operators $D_j$
             - are optimal if the cost functional
$ E = \|\rho \|^2 $ is at its minimum.

 {\bf Example:} Consider a twogrid iteration in which a standard
 relaxation sweep on $\L^0$ with iteration matrix $\rho_0$
                            is followed by exact solution of the
  coarse grid equation~(2.4). The second step leads to
 an updating with some iteration matrix $\rho_1$, and
 $\rho = \rho_0\rho_1$. Therefore one may estimate
                           $E\leq \|D_0\ \rho_0\|^2\ E_1 $ with
$E_1 = \|D_0^{-1}\ \rho_1\|^2 $.
 This form of the estimate~\cite{4} is motivated by the fact that
 the fine grid relaxation smoothens the error, but does not converge
 fast. Therefore $\|D_0\ \rho_0\|$ is suppressed, whereas $\|\rho_0\|$
 is not much smaller than $1$.

 Only $E_1 $ depends on the interpolation kernels etc. Therefore one
 can try to  optimize these parameters by minimizing $E_1$.

 Using the trace norm, $\|\rho \|^2 = {\rm tr}\ \rho \rho^{\ast }$, one finds
 \be
  E_1  =  {\rm Volume}^{-1}
          \sum_{z,w\in \L^0} |\Gamma_{zw}|^2 \ \ \mbox{with}
  \ \
  \Gamma  =  D_0^{-1} - \A^1\ D_1^{-1}\ C^1 \ . \label{Gamma} \ee
 Prescribing $C^1 $, and determining $D_1$ and $\A^1 $ by minimizing
 $E_1$ yields what we call the ``ideal interpolation kernel'' $\A^1$
 for a given restriction map $C^1$.
 Kalkreuter did twogrid iterations for~(2.1) in $4$ dimensions
 with $D_0$ as shown in section 6 below, using ideal
 interpolation kernels~\cite{12}.
 He found absence of critical slowing down
 for arbitrarily disordered $SU(2)$-gauge fields. This showed for
 the first time that multigrid could work in principle for arbitrarily
 strong disorder.
 The ideal kernel $\A^1_{zx}$
         is impractical for production runs, though. This is because it
 has exponential tails instead of vanishing for $z$ outside a
 neighbourhood of $x$.

 In the analytic multigrid approach to Euclidean
 quantum field theory~\cite{1},
 $\Gamma $ is known as the fluctuation field~ or high frequency field
 propagator. It has an infrared cutoff which makes it decay
 exponentially with distance $|z-w|$.  Typically, the stronger the
 decay, the smaller $E_1$.
 \addtocounter{section}{1}
 \setcounter{equation}{0}
 \setcounter{figure}{0}
 \setcounter{table}{0}

\vskipp{12}
 {\noindent \bf 5. Neural Multigrid}
\vskipp{5}

A feed-forward neural network~\cite{13} can perform the computations
to solve~(2.2)
by multigrid relaxation. The nodes of the network (``neurons'')
                                                  are identified
with points of the multigrid. There are two copies of the multigrid,
except that the last layer $\L^N $ is not duplicated.
In the standard multigrid approach, the basic interpolation kernels
$\A^j $ interpolate from one layer $\L^j $ of the multigrid to the
preceding one, $\L^{j-1}$. In this case the network looks like in
fig.~\ref{FIGUR1}.
\begin{figure}
\vspace{8cm}
\caption{A feed-forward neural multigrid architecture
(selected connections).} \label{FIGUR1}
\end{figure}
Each node is connected to some of the nodes in the preceding layer
in the neural network. In the upper half, the connection strength
{}from $x\in \L^j $ to $z\in \L^{j-1}$ is $\A^j_{zx}$. In the lower half,
node $z \in \L^{j-1} $ is connected to $x\in \L^j $ with strength
$R^j_{xz}$. In addition there is a connection of strength
$\omega_jd_{j,x}^{-1}$
($d_{j,x} \equiv (D_j)_{xx}$, $\omega_j \equiv$ damping parameter
on $\Lambda^j$)
between the two nodes
which represent the same
point $z$ in $\L^j$, $j<N$. These connections model the synapses in
a brain. According to Hebb's hypothesis of synaptical learning,
the brain learns by adjusting the strength of its synaptical connections.

            The network receives as input an approximate solution
$\xi $ of~(2.1), from which the residual $r^0=f^0-D_0\xi $
is then determined.                   It computes as output an improved
solution $O = \xi + \delta \xi $. The desired output (``target'') is
$\zeta = D_0^{-1}f^0$. $\delta \xi $ is a linear function of $r^0$.

Except on the bottom layer,
each neuron receives as input a weighted sum of the output of those
neurons below it in the diagram to which it is connected. The weights
are given by the connection strengths. Our neurons are linear because
our problem is linear. The output of each neuron
                                 is a linear function of the input.
(One may  take output $=$ input).

The result of the computation is
\be \delta\xi = ( \omega_0\ d_0^{-1}  +
\sum_{k\geq 1} \A^{[0k]}\ \omega_k\  d_k^{-1} \ R^{[k0]})\ r^0 \ee
where $R^{[k0]}=R^k...R^2R^1$ and $\A^{[0k]}= \A^1\A^2...\A^k$.

In principle, $R^j$ is determined by the restriction operators and by
the effective differential operator, $R^j
                   = C^j(1-\omega_{j-1}D_{j-1}d_{j-1}^{-1})$. But
actually, the restriction kernels $C^j $, effective differential
operators $D_j$ $(j>0)$ and their diagonal part $d_j $, and the damping
parameters $\omega_j$ $(j>0)$                          enter only
in the combination $R^j$ (assuming $\omega_jd_{j,x}^{-1}$ can be scaled
to $1$).                  They are therefore not needed
separately.

            The fundamental differential operator $D_0 $ and its
diagonal part $d_0$ are furnished as part of the problem.
The connection strengths (``synaptical strengths'') $\A^j_{zx}$,
$R^j_{xz}$ (and possibly the damping factor
                         $\omega_0$) need to be found by a learning
process in such a way that the actual output is as close as possible
to the desired output.

In supervised learning of a neural network~\cite{13}, a sequence
of pairs $(\xi^{\mu }, \zeta^{\mu })$ is presented to the network.
Given input $\xi^{\mu }$,
the actual output $O^{\mu }$ is compared to the target $\zeta^{\mu }$,
and the connection strengths are adjusted in such a way that the
cost functional
$$ E = \sum_{\mu } \|O^{\mu } - \zeta^{\mu } \|^2 $$
gets minimized. If  an iterative procedure to
achieve this minimization is specified,
                           one calls this a \it  learning rule. \rm

                Because of linearity, it suffices to consider the
equation $D_0\xi = f^0 $ in the limit of small $f^0$.
In the limit $f^0 \mapsto 0$, the target $\zeta^{\mu }=0$ for any input,
and $O^{\mu }= \rho\  \xi^{\mu }$ by~(4.1), with $\rho $ =
iteration matrix.

Taking for the sequence $\xi^{\mu }$ a complete orthonormal system
of functions on $\L^0$,
$$ E= \sum_{\mu } \|\rho\ \xi^{\mu }\|^2= {\rm tr}\ \rho \rho_{\ast }
\equiv \|\rho \|^2\ . $$
$E$ = minimum is the previous optimality condition of section 4 for
multigrid relaxation.

{\bf Conclusion:} \it Optimizing the kernels $\A $ and $R$ is a standard
learning problem for a feed forward neural net. \rm

 \addtocounter{section}{1}
 \setcounter{equation}{0}
 \setcounter{figure}{0}
 \setcounter{table}{0}

\vskipp{12}
 {\noindent \bf 6. Learning Rule Performance}
\vskipp{5}

The second variant, for which a learning rule was already described in
section 3, involves a slightly different neural net. Instead of the
connections between neighbouring layers of the multigrid, we have now
connections from $\L^0$ to $\L^k$ with connection strength
 $C^{[k0]}_{xz}$, and from $\L^j$ to $\L^0$ with connection strength
$\A^{[0j]}_{zx}$.  We adopt variational coarsening,
$    C^{[k0]} =  \A^{[0k]\ast} $. Then
                  all connection strengths are determined by
interpolation kernels $\A^{[0k]}$ which have to be learned.
The  damping factors $\omega_k$ were
set to $1$,
and $d_k$ is the   diagonal part of $D_k$ as before, with
\be  D_k =
                 \A^{[0k]\ast}D_0 \A^{[0k]}\ .  \ee

The learning rule of section 3 requires a process of ``hard thinking''
by the neural net. Neurons which have learned their lesson
already - i.e. which have their synaptical strengths fixed - are used to
instruct the rest of the neural net, adjusting the synaptical
strengths of the next multigrid layer of neurons.

A checkerboard
             variant of this algorithm was tested in 2 dimensions, using
SU(2)-gauge fields which were
                                 equilibrated with standard Wilson
action at various values of $\beta$, and
$D_0=-\Delta-\epsilon_0+\delta m^2 $.
                 $\epsilon_0$ is the lowest eigenvalue of the covariant
  Laplacian $-\Delta $, and $\delta m^2 > 0$.
 Conventional relaxation algorithms for solving~(2.1) suffer
{}from critical slowing down for such $D_0$, for any volume and
                                               small $\delta m^2$.

It turned out that it was not necessary to find accurate solutions
of the eigenvalue equation for the interpolation kernels $\A^{[0j]}$.
An approximation $\A^{[0j]}_{zx}$ to  $(-\Delta )^{-n} \delta_{zx}$ was
computed.
A total of four relaxation sweeps through each multigrid layer below $j$
($n=2$ inverse iteration steps \`{a} one V-cycle each) was enough.
The convergence rate of the following $\xi$-iteration is
shown in fig.~\ref{FIGUR2} for $\beta =1.0$.
\begin{figure}
\vspace{8cm}
\caption{Correlation time $\tau$ as function of the lowest eigenvalue
$\delta m^2$ in a representative gauge field configuration
equilibrated at $\beta = 1.0$. For the $64^2$ lattice, the correlation
time fluctutates very little with the gauge field configuration.}
\label{FIGUR2}
\end{figure}
The correlation time
$\tau $ is in units of MG-iterations. One MG-iteration involved one
sweep through each multigrid layer, starting with $j=0$.
Updating  $\xi $ at $x \in  \L^j$ changes $\xi $ by
$$ \delta \xi_z=  \A^{[0j]}_{zx}d_{j,x}^{-1}r^j_x\ ,  \ \
r^j=\A^{[0j]\ast }r^0 .$$
    Sweeps were actually performed in checkerboard fashion.
\clearpage
\subsection*{Acknowledgements}
We thank the HLRZ J\" ulich for computer time and assistance.

Support by Deutsche Forschungsgemeinschaft is gratefully
acknowledged.
%
%
%

%
\end{document}